\documentclass[twoside]{dis08}
\usepackage[latin1]{inputenc}
\usepackage[dvips]{graphicx,epsfig,color}
\usepackage{wrapfig,rotating}
\usepackage{amssymb,amsmath,array}

\begin{document}
\title{\raggedleft{\normalsize DESY 08-111} \\[0.5em]
Some News about Generalised Parton Distributions%
\,\thanks{Talk given at the 16th International Workshop on Deep Inelastic
    Scattering and Related Subjects, University College London, England,
    7--11 April 2008 \protect\cite{url}}
}
\author{Markus Diehl
\vspace{.3cm}\\
Deutsches Elektronen-Synchroton DESY \\
22603 Hamburg - Germany
}

\maketitle

\begin{abstract}
  I briefly discuss some recent developments (and recall some old news) in
  the theory and phenomenology of generalised parton distributions.
\end{abstract}


\section{Parametrising generalised parton distributions}

\begin{wrapfigure}{r}{0.47\columnwidth}
\centerline{\includegraphics[width=0.37\columnwidth]{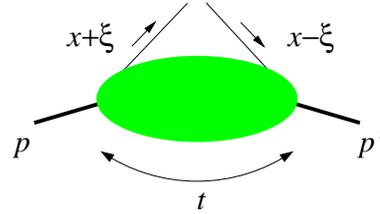}}
\caption{\label{fig:gpd} Basic variables describing a GPD.}
\end{wrapfigure}
As experimental data on deeply virtual Compton scattering (DVCS) and on
exclusive meson production is becoming more and more precise
\cite{exp-talks}, the demands on an adequate theory description are
rising.  An important part of any theoretical analysis using the framework
of generalised parton distributions (GPDs) is to devise a parametrisation
of these functions.  This is a complex task since GPDs depend on two
independent momentum fractions $x$ and $\xi$ and on the invariant momentum
transfer $t$ (see Fig.~\ref{fig:gpd}) and since the dependence on these
variables is subject to a number of consistency constraints.
Understanding this dependence is, however, not only a practical necessity
but can improve our very understanding of how partons are distributed in
the nucleon.

An important constraint on GPDs is the so-called polynomiality property:
the $n$th Mellin moment in $x$ is a polynomial in $\xi$ with degree $n$ or
$n-1$, depending on the particular GPD in question.  Being a direct
consequence of Lorentz invariance, this has turned out to be essential
even at the practical level, since it is required for the consistency of
dispersion relations for the processes where GPDs appear (see below).
Several methods that ensure this property in the construction of GPD
parametrisations have been used in phenomenology:
\begin{list}{$\bullet$}{\setlength{\leftmargin}{1.3em}%
    \setlength{\parsep}{0pt}}
\item the double distribution representation of GPDs.  The
  Radyushkin-Musatov ansatz \cite{Radyushkin:1998es,Musatov:1999xp} has
  been used in a large number of applications---including the VGG code
  \cite{Vanderhaeghen:1999xj}---but it should be emphasized that this is
  only one particular ansatz for the form of double distributions.  A
  Polyakov-Weiss $D$-term \cite{Polyakov:1999gs} is typically added to
  this ansatz so as to provide the term of order $\xi^{n}$ in the $n$th
  Mellin moment of the distributions $H$ and $E$.  In recent work
  \cite{Kumericki:2008di} the question was raised whether such a $D$-term
  can indeed be freely chosen or whether it is already fixed by the double
  distribution piece due to further consistency requirements.
  
  Recall also that there are several alternative double distribution
  representations of GPDs which generate the term of order $\xi^{n}$ in
  the Mellin moments by themselves and hence do not require a $D$-term
  from the start \cite{Teryaev:2001qm}.
\item the so-called dual parametrisation \cite{Guzey:2005ec} starts with a
  partial-wave decomposition in the $t$-channel and resums these partial
  waves into a series of two-variable functions (in contrast to the double
  distribution case, where one deals with a single three-variable
  function).  In practical applications, only the first or the first two
  functions in this series have been retained.

  Let me note that, in analogy to the magnetic and electric combinations
  of the Dirac and Pauli form factors, the $t$-channel partial waves are
  described by the combinations $H + E$ and $H + \frac{t}{4m^2} E$ of GPDs
  \cite{Diehl:2007jb}.  This feature is not incorporated in the
  parametrisations~\cite{Guzey:2005ec}.
\item an ansatz for the Gegenbauer, or conformal moments of the GPDs
  \cite{Kumericki:2008di}.  These moments depend on $\xi$, $t$ and a
  moment index $j$, which is to be considered as a variable in the complex
  plane in order to retrieve the GPDs.  This method generalises the
  representation of usual parton densities in terms of their Mellin
  moments, and it has the advantage that evolution and $\alpha_s$
  corrections can be implemented at NLO and partially at NNLO in a
  technically simple and numerically efficient way.
\end{list}

The lowest Mellin moments of the quark GPDs $H^q$ and $E^q$ are related to
the electromagnetic Dirac and Paul form factors, which are well measured
experimentally and are typically used as constraints in GPD
parametrisations.  Similarly, the axial form factor gives the lowest
moment of $\tilde{H}^u - \tilde{H}^d$.  Significant progress is being made
in calculating higher Mellin moments of quark GPDs on the lattice
\cite{Hagler:2007xi}.  Qualitative features found in these calculations
provide guidance for parametrising GPDs already now, and eventually one
can hope to use lattice results as a quantitative input.

An indirect constraint on the nucleon-helicity flip distribution $E^g$ for
gluons is provided by the sum rule $\smash{\int dx\, E^g + \sum_q \int
  dx\, x E^q = 0}$ at $t=0$ and $\xi=0$.  Lattice calculations give a very
small result for $\smash{\sum_q \int dx\, x E^q}$, with large
cancellations between $u$ and $d$ quark contributions.  This is in line
with the small value of the lowest moment $\sum_q \int dx\, E^q$ inferred
from the magnetic moments of proton and neutron.  Under the assumption
that $E^g$ does not change unusually fast when going to finite $t$ and
$\xi$, the above sum rule hence implies that the distribution $E^g$ is
itself small---in contrast to its counterpart $H^g{\mskip 1mu}$---unless
it has one or more nodes in $x$.  Current models
\cite{Diehl:2007hd,Goloskokov:2007nt} do not allow for such nodes (for
which there is no indication from data) and consequently have $E^g$ much
smaller than $E^u$ or $E^d$.  Going back to the point $t=0$ and $\xi=0$,
one should note that if $\int dx\, E^g$ is small then the total angular
momentum $\frac{1}{2} \int dx\, ( H^g + E^g)$ carried by gluons in the
nucleon according to Ji's sum rule is rather large, because $\int dx\, H^g
= \int dx\, x g(x)$ is about one half.

Further nontrivial constraints on GPDs are positivity conditions, which
ensure the interpretation of GPDs as densities or interference terms in
impact parameter space.  As in the case of the usual parton densities,
positivity conditions are preserved by leading-order evolution to higher
scales but need not hold at NLO or beyond.  It seems reasonable to demand
some explanation in cases where they break down (e.g.\ in the region of
very small $x$, where NLO corrections are known to become important).  The
most general form of the positivity conditions for GPDs is quite involved
\cite{Pobylitsa:2002iu}.  Simplified versions concern the skewness effect
and give e.g.\ an upper limit on $H^q(x,\xi,t)$ for $x>\xi$ in terms of
the geometric mean of the quark densities evaluated at
$(x\pm\xi)/(1\pm\xi)$ \cite{Pire:1998nw,Radyushkin:1998es}.  To implement
or even check these constraints in GPD parametrisations is difficult and
most often not done.  An exception are approaches based on calculating
double distributions in spectator models \cite{Pobylitsa:2002vw}, which
allow both polynomiality and positivity to be implemented automatically,
but so far have barely been used in phenomenological studies.  A different
type of simplified positivity condition involves the GPDs at $\xi=0$ and
limits for instance $E$ in terms of $H$ and $\tilde{H}$
\cite{Burkardt:2003ck}.  For simple analytic forms of the GPDs these
conditions can readily be checked---unfortunately this is not always done
in practice.

An approximate power-law dependence at small $x$, as suggested by simple
Regge phenomenology, works well in phenomenological parametrisations of
the usual parton densities, and it is natural to extend this to a
behaviour proportional to $x^{-(\alpha+\alpha' t)}$ for GPDs.  It is
important to note that DGLAP evolution changes the value of the effective
shrinkage parameter $\alpha'$ with the factorization scale
\cite{Diehl:2007zu}, just as it changes the value of the effective power
$\alpha$ at $t=0$.  This change can be rather slow, so that the mixing of
gluons and sea quarks under evolution does not by itself guarantee that
$\alpha'$ is similar for these partons at moderate scales
\cite{Diehl:2007zu}.


\section{From exclusive processes to GPDs}

It is well known that, to leading order in $\alpha_s$, the imaginary part
of the amplitude for DVCS or light meson production involves only GPDs at
the special points $x=\xi$ and $x=-\xi$.  While the impact parameter
representation is somewhat involved for GPDs at nonzero $\xi$
\cite{Diehl:2002he}, it simply gives the distance between the struck
parton and the \emph{spectator system} in this
case~\cite{Burkardt:2007sc}.

Whereas the factorization formula for the real part of the DVCS or meson
production amplitude requires the GPDs in the full $x$-region, a
representation based on dispersion relations \cite{Teryaev:2005uj}
involves, at leading order, only the GPDs at $|x| =\xi$ plus a subtraction
term which depends on the coefficients of $\xi^{n}$ in $\int dx\, x^{n-1}
H(x,\xi,t)$ and can hence be expressed through the $D$-term in the
Polyakov-Weiss representation.  This implies that at tree-level accuracy,
exclusive processes are only sensitive to GPDs at $|x| =\xi$ and to one
$t$-dependent subtraction term for each parton species.  Both the
evolution of the GPDs and explicit NLO corrections to the amplitude
involve the region $|x| \ge\xi$, in addition to a more complicated
subtraction constant \cite{Diehl:2007jb,Kumericki:2008di}.  To access this
information experimentally, one must be sensitive to violations of Bjorken
scaling and thus needs a sufficient lever arm in $Q^2$ at given $x_B$.

The polynomiality of Mellin moments, which ties together the GPDs at $|x|
< \xi$ and at $|x| \ge\xi$, is crucial for the consistency of the
dispersion relations just mentioned.  In \cite{Kumericki:2008di} an
explicit construction is given that allows one to reconstruct a GPD in the
region $|x| < \xi$ from its knowledge at $|x| \ge\xi$, apart from a
possible $D$-term ambiguity.

Deeply virtual Compton scattering not only offers a large number of
observables that can be evaluated in the GPD framework
\cite{Belitsky:2001ns} but is also under very good theoretical control,
with radiative corrections known at NLO and in part at NNLO.  In numerical
studies these corrections were found to be of moderate size
\cite{Mueller:2005nz}, except for important effects in scaling violation
at small $x$, which are of the same nature as those in inclusive DIS and
stem from the behavior of the singlet evolution kernels.  Nevertheless,
NLO corrections of the order of $20\%$ are not uncommon in collider or
fixed-target kinematics.  While a leading-order analysis of DVCS data
should be adequate to reveal basic features and to provide a starting
point, I see no good reason why analyses should be limited to leading
order (including the neglect of evolution) as experimental data is
becoming more precise.  The NLO scattering amplitudes for DVCS are
reasonably simple for practical use, and the LO evolution of GPDs has been
implemented in a fast numerical code \cite{Vinnikov:2006xw}.

Meson production is harder to analyse quantitatively, as both power
corrections \cite{Goloskokov:2007nt} and higher orders in $\alpha_s$
\cite{Belitsky:2001nq,Diehl:2007hd} are larger than for DVCS according to
the estimates in the literature.  The often substantial size of NLO
corrections should not surprise us, given that $\alpha_s \sim 0.4$ for
scales relevant to a large part of the experimental data.  Cross section
ratios are often assumed to be less sensitive to corrections.  The study
\cite{Diehl:2007hd} found that this is sometimes true but not always.  The
ratio of $\omega$ and $\rho$ production cross sections at moderate $x_B$
is indeed quite stable w.r.t.\ NLO corrections.  For the cross section
ratio of $\phi$ and $\rho$ production this is, however not the case, since
the former is dominated by gluon exchange and the latter by a mixture of
quark and gluon exchange, which are affected in different ways by NLO
corrections.  For the GPD models considered in \cite{Diehl:2007hd}, the
transverse target spin asymmetries for $\rho$, $\omega$, and $\phi$
production are very small at tree level, due to cancellations between
different contributions and to a small relative phase between the
amplitudes that have to interfere in order to make the spin asymmetry
nonzero.  The NLO corrections on these small asymmetries were found to be
large.  A careful case-by-case analysis is hence required before one can
assert the stability of cross section ratios or asymmetries.


\section*{Acknowledgments}

It is a pleasure to thank R.~Devenish, M.~Wing, and their co-organisers
for hosting a great conference.


\begin{footnotesize}

\end{footnotesize}


\end{document}